\documentstyle[aps,epsfig,prb]{revtex}              

\def\bk{ {\bf k} }
\def\bq{ {\bf q} }
\def\tveps{ {\tilde\epsilon} }

\def\im{ \,{\rm Im}\, }
\def\re{ \,{\rm Re}\, }

\begin{document}
\twocolumn[\hsize\textwidth\columnwidth\hsize\csname
@twocolumnfalse\endcsname               
\draft

\title{Electron-Phonon Interaction and Ultrasonic Attenuation in the
Ruthenate and Cuprate superconductors}
\author{M. B. Walker, M. F. Smith, and K. V. Samokhin}
\address{Department of Physics,
University of Toronto,
Toronto, Ont. M5S 1A7 }
\date{\today}
\maketitle

\widetext                   %
\begin{abstract}
This article derives an electron-phonon interaction suitable for
interpreting ultrasonic attenuation measurements in the ruthenate and
cuprate superconductors.  The huge anisotropy found experimentally
(Lupien {\em et al}., 2001) in Sr$_2$RuO$_4$ in the normal state is
accounted for in terms of the layered square-lattice structure
of Sr$_2$RuO$_4$, and the dominant contribution to the attenuation in
Sr$_2$RuO$_4$ is found to be due to electrons in the $\gamma$ band.
The experimental data in the superconducting state is found to be
inconsistent with vertical lines nodes in the gap in either (100) or
(110) planes.  Also, a general method, based on the use of symmetry,
is developed to allow for the analysis of ultrasonic attenuation
experiments in superconductors in which the electronic band structure
is complicated or not known. Our results, both for the normal-state anisotropy, and
relating to the positions of the gap nodes in the superconducting
state, are different from those obtained from analyses using a more
traditional model for the electron-phonon interaction in terms of an
isotropic electron stress tensor. Also, a brief discussion of the ultrasonic
attenuation in UPt$_3$ is given.
\end{abstract}

\pacs{PACS numbers: 74.70.Pd, 74.20.Rp, 74.25.Ld}

\vfill      
\narrowtext         %

\vskip2pc]  

\section{Introduction}
\begin{figure}
\centerline{\epsfig{file=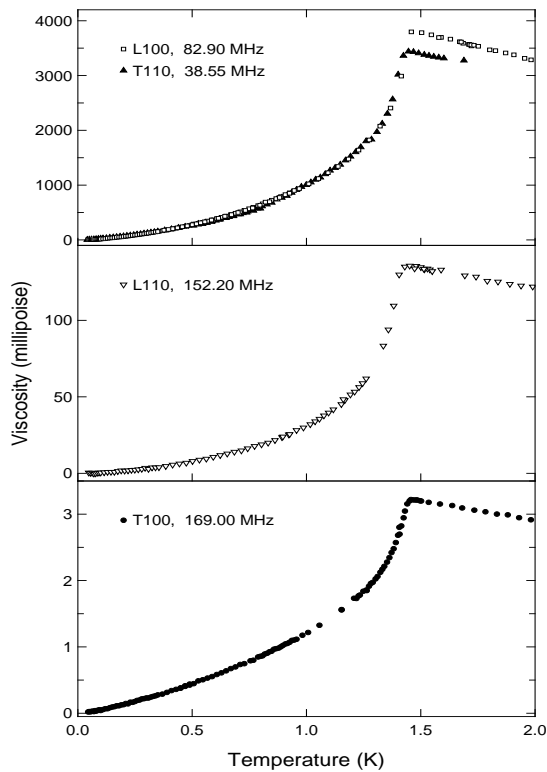,height=4in,width=2.8in}}
\vspace{10pt}
\caption{Experimental data on the mode viscosity for the four
in-plane sound wave modes taken from Ref.\
\protect\onlinecite{lup01}.  The mode viscosity $\eta$ plotted here
is related to the attenuation $\alpha$ by the formula $\eta = \alpha
\rho v_s^3/(2\pi \nu)^2$, where $\rho$ is the density, $v_s$ is the
sound velocity and $\nu$ is the frequency.  All sound wave modes have
both the direction of propagation and the polarization lying in the
basal plane.  The nature of the mode ($L$ for longitudinal and $T$ for
transverse) as well as the Miller indices of the propagation
direction are shown in the figure.}
\label{Sr_expt}
\end{figure}

This article describes a theory of ultrasonic attenuation in the
ruthenate and cuprate superconductors.  It has been stimulated
largely by the work of Lupien et al
\protect\cite{lup01} which presented the results of detailed
experimental measurements of ultrasonic attenuation in both the normal
and superconducting states of Sr$_2$RuO$_4$.  The goal of that work
was to use ultrasonic attenuation as a tool to gain information on
the presence and location of nodes in the superconducting gap.  Quite
unexpectedly, however, they found a huge anisotropy in the measured
attenuation, even in the normal state. Some of their results are
shown in Fig.\ \ref{Sr_expt}. Notice that the normal-state
attenuation of the longitudinal wave propagating along the [110]
direction is lower by a factor of approximately 30 than that of the
[100] longitudinal wave.  Furthermore, the attenuation of the
transverse wave along the [100] direction is lower than that of the
transverse wave along the [110] direction by a factor of more than
1000.  Lupien et al. note that their results indicate the need for a
new theory of the electron-phonon interaction allowing for a
significant variation for the different sound wave modes. They also
suggest that the lack of a reliable theory of ultrasonic attenuation
in Sr$_2$RuO$_4$ has greatly hindered the use of this technique as a
tool for gaining information about the location of the gap nodes in
this material.

The use of ultrasonic attenuation as a tool to locate the positions
of nodes in the energy gaps of superconductors has been described,
for example, by Moreno and Coleman \protect\cite{mor96}.  In their
model, the phonon strain field couples to a stress tensor describing
the flow of electron momentum.  The electron stress tensor is taken
to be that appropriate to an isotropic electron fluid.  This
approach, which can be traced back to early work on ultrasonic
\noindent attenuation in heavy-fermion superconductors
\protect\cite{hir86,sch86,var86}, in $s$-wave superconductors \protect\cite{tsu61,kad64} and in normal
metals \protect\cite{blo59}, has also been used in the most recent
attempts to understand ultrasonic attenuation in Sr$_2$RuO$_4$
\protect\cite{gra00,wu01}. The use of an electron stress tensor
appropriate to an isotropic fluid gives an elegant and simple
formulation of the theory.  However, it seems clear that a new
approach is needed if one wishes to be able to account in detail for
the ultrasonic attenuation observed in the recent experiments on
Sr$_2$RuO$_4$.  The approach developed in this article takes account
of the crystalline and electronic structure of Sr$_2$RuO$_4$, but also
can be shown to reduce to the traditional isotropic electron-stress-tensor
approach in an appropriate limit (this limit not being appropriate
for Sr$_2$RuO$_4$).

This article shows that the strong anisotropy of the ultrasonic attenuation
in Sr$_2$RuO$_4$ is intimately connected with its 
layered square-lattice structure.
We start from the idea that the electronic structure
of Sr$_2$RuO$_4$ in the neighborhood of the Fermi surface can be well
described in terms of a simple tight-binding Hamiltonian,
\protect\cite{agt97,maz97} in which a central role is played by
hopping matrix elements describing the hopping of an electron from
one ruthenium ion to a nearest-neighbor or next-nearest-neighbor
ruthenium ion.  To develop a model for the electron-phonon
interaction, we assume that the hopping matrix element depends on the
distance between the ions, so that if a sound wave stretches the
distance between two ions, the matrix element describing the hopping
of an electron between these two ions changes.  It is easily seen
that a transversely polarized sound wave travelling in the [100]
direction in the ruthenium-ion square lattice does not stretch
the nearest-neighbor bond between two
ruthenium ions, and hence is not coupled to the electrons (at least
by a nearest-neighbor coupling).  This is the reason for the
extremely low attenuation of the T100 sound wave reported in Ref.\
\protect\onlinecite{lup01} (see Fig.\ \ref{Sr_expt}).  It is also
easily seen that for a two- or three-dimensional hexagonal lattice,
or for a three-dimensional body-centered or face-centered
lattice, or if the next-nearest-neighbor interaction is large, this
argument does not apply.  Thus the highly unusual strongly anisotropic
ultrasonic attenuation observed in Sr$_2$RuO$_4$ is directly related to
the layered square-lattice structure of this material. The anisotropy
in the attenuation of longitudinal waves (e.g. for the L100 and L110
waves of Fig.\ \ref{Sr_expt}) also finds an elementary qualitative
and quantitative explanation in terms of the model developed in
detail below: this explanation depends on the details of the Fermi
surface geometry and will be given later.

After developing a detailed model for the electron-phonon interaction
in Sr$_2$RuO$_4$, and
confirming that the model accounts well for the ultrasonic
attenuation in the normal state, we proceed to an analysis of the
ultrasonic attenuation in the superconducting state with the
objective of gaining information on the positions of nodes in the
superconducting energy gap.
The basic ideas here have been clearly set out in the article by
Moreno and Coleman \protect\cite{mor96}.  They show that for a given
sound wave propagation direction and polarization, the nodes in the
gaps can be described as either ``active'' or ``inactive'',  and that
the temperature dependence of the ultrasonic attenuation is very
different in the two cases.  These ideas are described in greater
detail below, and exploited to gain information on the positions of
the gap nodes in Sr$_2$RuO$_4$. Our calculations are done in the
hydrodynamic limit. The opposite (``quantum'') limit relevant for either
very pure samples, or very high frequencies, has been considered
in Ref.\ \protect\onlinecite{vekh99}.

A consequence of our work is the recognition that the traditional
isotropic electron-stress-tensor model of the electron-phonon interaction
makes certain nodes ``accidentally'' inactive for the case of
longitudinal sound waves.  This accidental inactivity is a
consequence of the fact that the isotropic electron-stress-tensor model of
the electron-phonon interaction is not sufficiently general to
describe realistically the electron-phonon interaction.  In a more
realistic model, all nodes are active for longitudinal phonons.
Thus, the isotropic electron-stress-tensor model for the electron-phonon model
gives misleading results for the temperature dependence of the
ultrasonic attenuation associated with longitudinal phonon modes in
some cases, and should not be used in studies aimed at determining
the positions of the nodes in the gap of an unconventional
superconductor.

Although the emphasis in this article will be on ultrasonic
attenuation in Sr$_2$RuO$_4$ because of the availability of
experimental data for this material \protect\cite{lup01}, it should
be noted that our detailed results for the $\gamma$ band are also
applicable (except for the interlayer interaction) to cuprate
superconductors such as YBa$_2$Cu$_3$O$_{6+x}$; furthermore it is a
simple matter to develop an appropriate interlayer electron-phonon
interaction for the YBa$_2$Cu$_3$O$_{6+x}$ structure analogous to the
result for Sr$_2$RuO$_4$ developed here.
The analysis in Sec. \ref{symmetry_arguments} allows us to
make some general statements about the symmetry-imposed properties
of the electron-phonon interaction  and their manifestations
for the ultrasonic attenuation in unconventional superconductors,
in particular, in UPt$_3$.

The discovery \protect\cite{mae94} of superconductivity in the
layered perovskite Sr$_2$RuO$_4$, and the proposal
\protect\cite{ric95} that the superconducting Cooper pairs in that
material formed in a spin-triplet state has stimulated considerable
interest and study (see Ref.\ \protect\onlinecite{mae01} for a review
of the properties of Sr$_2$RuO$_4$, and Ref.\ \protect\onlinecite{Book}
for a review of the symmetry classification and the physical properties of
unconventional superconductors in general).
Until recently, it was thought that the superconducting order in Sr$_2$RuO$_4$
could be described by the order parameter
${\bf d}({\bf k}) \propto {\bf \hat{z}}(k_x + ik_y)$
\protect\cite{agt97,ric95}.  Because the Fermi surface of
Sr$_2$RuO$_4$ has a quasi-two-dimensional cylindrical form
\protect\cite{mac96}  with no points at $k_x = k_y = 0$, the Fermi
surface for this order parameter is fully gapped.
More recently, however, power-law temperature dependence more
characteristic of a gap having line nodes have been found in a
number of experiments, including specific heat
\protect\cite{nis99,nis00}, NQR \protect\cite{ish00}, penetration
depth \protect\cite{bon00}, thermal conductivity
\protect\cite{tan01,iza01} and ultrasonic attenuation
\protect\cite{lup01}. It is not easy to reconcile the presence of
triplet Cooper pairs, a broken-time-reversal symmetry, and line nodes
in the gap.  A number of proposals to do so have nevertheless been
made. These include spin-triplet states characterized by vectors
${\bf d}({\bf k})$ of the form ${\bf d}({\bf k}) \propto {\bf {\hat{z}}}
(\sin(k_x a) + i \sin(k_y a))\;$ \protect\cite{miy99,dah00,kuw00} and
having vertical line nodes in (100) planes,
so-called $f$-wave states characterized by ${\bf d}({\bf k}) \propto
{\bf {\hat{z}}} (k_x + i k_y) k_x k_y$
or by ${\bf d}({\bf k}) \propto {\bf {\hat{z}}} (k_x + i k_y) (k_x^2
- k_y^2)\;$ \protect\cite{gra00,wu01,dah00,has00}, having vertical
line nodes in (100) and (110) planes, respectively,
by $f$-wave states characterized by ${\bf d}({\bf k}) \propto {\bf
{\hat{z}}} k_z(k_x + i k_y)^2\;$ \protect\cite{won00,mak99} and having
horizontal lines nodes in the plane $k_z = 0$, by states
characterized by ${\bf d}({\bf k}) \propto {\bf {\hat{z}}} (k_x + i
k_y) \cos(k_z c)\;$ \protect\cite{has00} and having horizontal line
nodes in the plane $k_z = \pi/(2c)$, or by states characterized by
${\bf d}({\bf k}) \propto {\bf {\hat{z}}} (\sin(k_xa/2)\cos(k_ya/2) +
i \cos(k_xa/2)\sin(k_ya/2))\cos(k_zc/2)\;$ \protect\cite{has00,zhi01}
and having horizontal line nodes in the plane $k_z = \pi/c$.  The
analysis of the ultrasonic attenuation experiments given below allows
many of these possibilities to be ruled out, and suggests that attention
be focused on those possibilities characterized by the existence of
horizontal line nodes.

The structure of the article is as follows.
Section \ref{model_formulation} develops a tight-binding model
of the electron-phonon interaction accounting for the details of
the layered square-lattice structure occuring in the ruthenate
and cuprate superconductors. Section \ref{normal_state} evaluates a
formula giving the ultrasonic attenuation in the normal state of
Sr$_2$RuO$_4$ in terms of the model electron-phonon interaction
given in Section \ref{model_formulation}, and shows that the
extremely strong and unusual anisotropy of the attenuation is
accounted for by the model.  Section \ref{uss} makes use of the
model electron-phonon interaction to determine the activity or
inactivity of the gap nodes for various proposed superconducting
gap structures for Sr$_2$RuO$_4$ thus allowing statements to
be made about which of the various proposals for the gap structure
 are consistent
with the experimental ultrasonic attenuation data.  Section
\ref{symmetry_arguments} shows that a detailed model of the
electron-phonon interaction is not necessary to determine which
nodes in the superconducting state are active or inactive, by
showing how to obtain such information from symmetry arguments
only.  Such arguments are sufficiently powerful to be applicable
to cases where the development of a detailed model of the
electron-phonon interaction is not available.  Our detailed
model for Sr$_2$RuO$_4$ is shown to be consistent with these
arguments (although the isotropic electron stress tensor model
is not), and furthermore, some new results relating to the
interpretation of ultrasonic attenuation in UPt$_3$ are
presented.  The Appendix gives a discussion of the universality
of the low-temperature, temperature-independent contribution to
the attenuation in the superconducting state, showing that the
presence (or absence) of universal behavior is associated with
the activity (or inactivity) of the gap nodes.

\section{Electron-Phonon Interaction in Layered Cuprate and Ruthenate
Superconductors}
\label{model_formulation}

\begin{figure}
\centerline{\epsfig{file=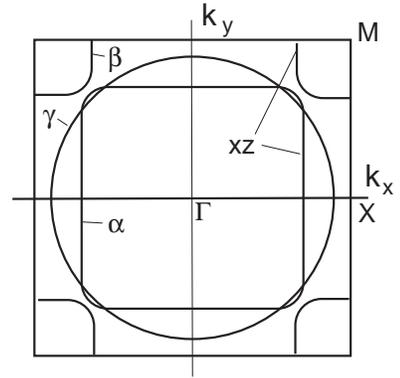,height=2in,width=2in}}
\vspace{10pt}
\caption{Schematic of the Fermi surface of Sr$_2$Ru$_2$O$_4$ showing the
$\alpha$, $\beta$ and $\gamma$ bands.  Also shown are portions of the $\alpha$
and $\beta$ bands that are predominantly $xz$ in character.}
\label{FS}
\end{figure}

The approximate two-dimensional nature \protect\cite{mac96} of the
Fermi surface of Sr$_2$RuO$_4$ suggests that it can be described, to
a first approximation, in terms of electrons interacting principally
through intraplanar interactions.  The Fermi surface consists of
three sheets (see Fig.\ \ref{FS}), which can be thought of as being
derived from the three ruthenium orbitals, $d_{xy}, d_{xz},$ and
$d_{yz}$ \protect\cite{mae01,ogu95,sin95}.  The Hamiltonian
describing the band structure of a single plane can thus be written
\protect\cite{agt97}
\begin{equation}
     H_{plane} = \sum_{\nu,\nu^\prime,{\bf n,n}^\prime,\sigma}
         t_{\nu,\nu^\prime}({\bf r_n} - {\bf r}_{{\bf n}^\prime})
         c_{\nu,{\bf n},\sigma}^\dagger
         c_{\nu^\prime,{\bf n}^\prime,\sigma},
     \label{H}
\end{equation}
where  $c_{\nu,{\bf n},\sigma}^\dagger$   creates an electron with
spin $\sigma$ in the $\nu$th ($\nu = \{xy, xz, yz\}$) ruthenium
orbital on the ruthenium ion at the site ${\bf n}$ of the ruthenium
square lattice.
\noindent Because of the $\sigma_z$ plane of reflection
symmetry at the center of the RuO$_4$ plane, there is no overlap
between the $xy$ orbitals and those of $xz$ and $yz$ symmetry.  This
means that one sheet of the Fermi surface (the $\gamma$ sheet)  can
be attributed to the $xy$ orbitals, while the $\alpha$ and $\beta$
sheets result from the hybridization of the $xz$ and $yz$ orbitals.
Also, from symmetry, there is no nearest-neighbor overlap between the
$xz$ and $yz$ orbitals, while the largest overlap integral for an
$xz$ orbital is expected to be with nearest-neighbor ions in the $\pm
x$ directions, since the lobes of these orbitals point in these
directions rather than in the $\pm y$ directions.  This means that the
$xz$ and $yz$ orbitals form approximately independent one-dimensional
bands, with the hybridization of these band giving relatively small
perturbations to the energies, except where the bands cross.  A
schematic view of the Fermi surface is shown in Fig.~\ref{FS}.

To derive an expression for the electron-phonon interaction, we
assume that the hopping matrix elements in Eq. (\ref{H}) in the
vibrating lattice depend on the
instantaneous positions of the ruthenium ions,
${\bf r_n} = {\bf r}^{(0)}_{\bf n}+{\bf u_n}$, where
${\bf r}^{(0)}_{\bf n}$ is the equilibrium position of the ion at lattice site
${\bf n}$, and ${\bf u_n}$ is its displacement from equilibrium.  The
lowest order contribution to the electron-phonon interaction is found
by expanding the hopping matrix elements in powers of the ionic
displacements ${\bf u_n}$ and keeping only linear terms.  First
consider doing this for only the nearest-neighbor hopping matrix
elements for the $xz$ orbitals that lie in the $\pm x$ directions
relative to each other, since this is a relatively simple effectively
one-dimensional case. This results in the interaction
\begin{eqnarray}
     H_{e-ph, plane}^{xz} &=& \frac{g^{xz}}{a} \sum_{{\bf n},\sigma}
     (u_{{\bf n},x} - u_{{\bf n+a},x})
     \nonumber \\
     & &  \times (c_{xz,{\bf n},\sigma}^\dagger c_{xz,{\bf n+a},\sigma}
     + c_{xz,{\bf n+a},\sigma}^\dagger c_{xz,{\bf n},\sigma}).
     \label{He-ph,plane}
\end{eqnarray}
Here $u_{{\bf n},x}$ is the $x$th component of displacement of the
${\bf n}$th ion in the plane, and $u_{{\bf n+a},x}$ refers to the ion
one Bravais-lattice vector ${\bf a}$ in the positive $x$ direction
from the ${\bf n}$th ion.  Notice that, for reasons of symmetry, this
expression does not contain displacements normal to the bond axis.
Thus (even though we have not assumed that the forces between the
ions are central), only displacements that stretch the bond distance
contribute to the electron-phonon interaction in this case.
Introducing the lattice Fourier transforms of the site variables in
Eq. (\ref{He-ph,plane}), and summing this Hamiltonian over all planes
in the crystal, gives
\begin{equation}
     H_{e-ph}^\nu = \frac{1}{\sqrt{N}}
  \sum_{{\bf q},j,{\bf k},\sigma,{\bf G}}
             g_{{\bf k}+\frac{1}{2}{\bf q,q},j}^\nu A_{{\bf q},j}
             c_{\nu,{\bf k+q+G},\sigma}^\dagger c_{\nu,{\bf k},\sigma},
     \label{He-ph}
\end{equation}
where $\nu = xz$, ${\bf G}$ is a reciprocal lattice vector, $A_{{\bf q},j} =
a_{-{\bf q},j}^\dagger + a_{{\bf q},j}$,
$a_{{\bf q},j}$ destroys a phonon corresponding to wave vector ${\bf
q}$ and polarization $j$, and $N$ is the total number of ruthenium
ions in the crystal. The wave vectors $\bq$ and $\bk$ are
three dimensional wave vectors and this Hamiltonian is for the whole
crystal.  Here
\begin{equation}
   g_{{\bf k}+\frac{1}{2}{\bf q,q},j}^\nu =
   g_{{\bf q},j}F^\nu_j(\bk,\bq),
     \label{g_total}
\end{equation}
with
\begin{eqnarray}
     g_{{\bf q},j} & = & -\sqrt{2} i \left(\frac{\hbar
    \omega_{{\bf q},j}}{M v_j^2} \right)^\frac{1}{2},
   \label{g} \\
     F^\nu_j(\bk,\bq) & = & g^\nu \sum_{\bf R}
     (\hat{\bf q}\cdot\hat{\bf R})(\hat{\bf R}\cdot {\bf e}_j({\bf q}))
    \cos(\bf{k \cdot R}), \label{F}
\end{eqnarray}
where $\omega_{{\bf q},j}$ is a phonon frequency, $v_j$ is the
sound velocity for a phonon of polarization $j$, ${\bf e}_j(\bq)$ is
a unit vector in the direction of the phonon polarization, $M$ is the
total mass of the ions in a primitive unit cell, a hat (as in
$\hat{\bf q}$) indicates a unit vector.
Since we are interested
in this article only in low frequency phonons, the expression
(\ref{F}) is given to the lowest non trivial order in the phonon wave vector
${\bf q}$.  The result of this section for the interaction of phonons
with the $xz$ electrons as described by Eq. (\ref{He-ph,plane}), is
given by Eqs. (\ref{He-ph}), (\ref{g_total}), (\ref{g}), and (\ref{F}) with
$\nu = xz$ and with the sum over ${\bf R}$ containing a single term
with ${\bf R}$ equal to the Bravais lattice vector ${\bf a}$.

Similarly, the interaction of phonons with the $yz$ electrons is
described by Eqs. (\ref{He-ph}), (\ref{g_total}), (\ref{g}), and (\ref{F})
with $\nu = yz$ and with the sum over ${\bf R}$ containing a single
term with ${\bf R}$ equal to the Bravais lattice vector ${\bf b}$.
Also, the coupling constant is called
$g^{\alpha \beta} \equiv g^{xz} = g^{yz}$.

The nearest-neighbor interaction of phonons with the $xy$ ($\gamma$
band) electrons is described by Eqs. (\ref{He-ph}), (\ref{g_total}),
(\ref{g}), and (\ref{F}) with $\nu = xy$ and with the sum over ${\bf R}$
containing two terms with ${\bf R}$ equal to the Bravais lattice
vectors ${\bf a}$ and ${\bf b}$.  The coupling constant is called
$g^{\gamma}$.

The next-nearest-neighbor interaction of phonons with the $xy$
($\gamma$ band) electrons is described by Eqs. (\ref{He-ph}),
(\ref{g_total}), (\ref{g}), and (\ref{F}) with $\nu = xy$ and with the sum
over ${\bf R}$ containing two terms with ${\bf R}$ equal to the
Bravais lattice vectors ${\bf a+b}$ and ${\bf a-b}$.  The coupling
constant is called $g'{}^\gamma$.

So far, we have considered only electron-phonon interactions
associated with the stretching of bonds between ions lying in a
single plane.  It is reasonable to expect these interactions to be
larger than the interplanar ones because the separation between
RuO$_4$ planes is relatively large, although this should be confirmed
by comparison with experiment.  However, the intraplanar interactions
that we have considered so far do not affect transverse phonon modes
that have their wave vector in the basal plane and their polarization
perpendicular to this plane.  This is because, to first order in the
displacements, such phonon modes do not stretch the bond distances,
and more rigorously, because the basal plane is a plane of reflection
symmetry.  Hence, to account for the attenuation of these phonon
modes, it is necessary to consider interplanar interactions.

In describing the interplanar interactions it is reasonable to
consider first interactions involving $xz$ and $yz$ orbitals since
these orbitals have lobes sticking out of the plane, whereas the
lobes of the $xy$ orbitals lie in the plane. The fact that the parts
of the Fermi surface that are made up from the
$xz$ and $yz$ orbitals show the largest corrugations
\protect\cite{ber00} along the $c$-axis direction support this
consideration. \protect\cite{zhi01} In addition we choose to consider interactions between
ruthenium ions at the corner and the body center of the unit cell
(which are the nearest-neighbor interplane pairs). Unfortunately, the
$xz$ and $yz$ orbitals are not a convenient basis for describing this
interaction. For this reason, the new basis, $c_{\xi z} = (c_{xz} +
c_{yz})/\sqrt{2}$ and $c_{\eta z} = (-c_{xz} + c_{yz})/\sqrt{2}$ will
be introduced.  Clearly, the $\xi z$ orbitals will interact well with
themselves along body diagonals in the directions of the vectors
${\bf a+b+c}$ and ${\bf -a-b+c}$, while the $\eta z$ orbitals will
interact well with each other along the body diagonals ${\bf -a+b+c}$
and ${\bf a-b+c}$.

The nearest-neighbor body-diagonal interactions between $\xi z$ orbitals are
described by Eqs. (\ref{He-ph}), (\ref{g_total}), (\ref{g}), and (\ref{F})
with $\nu = \xi z$ and with the sum over ${\bf R}$ containing two
terms with ${\bf d}$ equal to the Bravais lattice vectors
$\frac{1}{2}({\bf a+b+c})$ and $\frac{1}{2}({\bf -a-b+c})$.   Also,
the nearest-neighbor body-diagonal interactions between $\eta z$
orbitals are described by Eqs. (\ref{He-ph}), (\ref{g_total}), (\ref{g}),
and (\ref{F}) with $\nu = \eta z$ and with the sum over ${\bf R}$
containing two terms with ${\bf R}$ equal to the Bravais lattice
vectors $\frac{1}{2}({\bf -a+b+c})$ and $\frac{1}{2}({\bf a-b+c})$.
The
coupling constants for these two interactions satisfy $g^{\xi z} = g^{\eta z}$.
Because the best simple approximation to the band structure is in terms of the
$xz$ and $yz$ states, and not in terms of the $\xi z$ and $\eta z$
states, the electron-phonon interaction derived in the $\xi z$ and
$\eta z$ representation should be transformed back to the $xz$ and
$yz$ representation.

\section{Ultrasonic Attenuation in the normal state of Sr$_2$RuO$_4$}
\label{normal_state}

As noted in the Introduction, there is an extremely strong anisotropy
of the ultrasonic attenuation in the normal state of Sr$_2$RuO$_4$.
This section shows how this anisotropy can be accounted for in terms
of the electron-phonon interaction just described, and the known
Fermi-surface geometry of Sr$_2$RuO$_4$.

The attenuation constant $\alpha_j(\bf q)$ for an acoustic phonon
of wave vector ${\bf q}$ and polarization $j$ is given by
$\alpha_j({\bf q}) = (v_j\tau_{{\bf q},j})^{-1}$, where
$\tau_{{\bf q},j}$ is the phonon lifetime.
Because the formula for the ultrasonic attenuation in the
superconducting state is closely related to that for the normal
state, the latter more general formula for the superconducting state
will be given here, and the formula for the attenuation in the normal
state immediately follows. The result of evaluating the phonon
lifetime in the hydrodynamic limit (the electron quasiparticle mean
free path much shorter than the phonon wavelength) from the
self-energy of phonon Green's function in the superconducting state is
\begin{eqnarray}
    \frac{1}{\tau_{{\bf q},j}} & = &
    \frac{16\omega_{{\bf q},j}^2 N_F}{\rho v_j^2}
    \int_0^\infty \frac{d\epsilon}{\epsilon}
    \left( -\frac{\partial f}{\partial \epsilon} \right) \nonumber \\
    & & \times\left\langle F^2_j({\bf k,q})\tau_{\bf k}
    \re \sqrt{\epsilon^2 - |\Delta_{\bf k}|^2} \right\rangle_{FS},
    \label{tau_s}
\end{eqnarray}
where $N_F$ is the density of states at the Fermi level in the normal state,
$f(\epsilon)$ is the Fermi distribution function, and the Fermi surface average
is defined by
\begin{equation}
     \left\langle F^2_j(\bk,\bq) \right\rangle_{FS} =
     \frac{ \int F^2_j(\bk,\bq) dS_{\bf k}/v_{\bf k} }
         {\int  dS_{\bf k}/v_{\bf k}}.
     \label{FSavg}
\end{equation}
The integration over $dS_{\bf k}$ in this equation is over all (the $\gamma$, the $xz$ and the $yz$) sheets of the Fermi surface, with the electron-phonon matrix element chosen appropriately for each sheet.
The expression (\ref{tau_s}) is valid for all singlet superconducting phases,
as well as for unitary triplet phases, for which $|\Delta_{\bf k}|^2=
|{\bf d}({\bf k})|^2$.
This formula is similar to that employed in Ref.\
\protect\onlinecite{mor96}, except that in our expression the
electron-phonon matrix element $F_j(\bk,\bq)$  replaces the isotropric electron
stress tensor of Ref.\ \protect\onlinecite{mor96}.  Also, the formula has  been generalized to be applicable to anisotropic multi-sheet
Fermi surfaces.
The quantity $\tau_{\bk}$ is the Bogoliubov quasiparticle lifetime, and
for Eq. (\ref{tau_s}) to
be valid, the condition $k_B T \gg \hbar / \tau_{\bf k}$ must be
satisfied (see Appendix for a more detailed discussion).

One aspect of the above discussion that is unsatisfactory is the failure 
to give a full treatment of the Coulomb interaction so that charge 
neutrality is preserved in the distorted lattice that occurs in the 
presence of a longitudinal sound wave.  A microscopic treatment of this 
question for our tight-binding multi-band model
is beyond the scope of this article.  This question has however been treated 
in early work on  ultrasonic attenuation in metals with anisotropic 
Fermi surfaces \protect\cite{akh57}. There it was found that  charge 
neutrality can be simply imposed in terms of an appropriately chosen 
spatially varying chemical potential. This leads to the same formula 
for the attenuation as is obtained when the charge-neutrality correction 
is neglected, except that the original electron-phonon matrix element 
is replaced by a related effective electron-phonon matrix element.  
Translated into our notation, their result is that the electron-phonon 
matrix element $F_j(\bk,\bq)$ in Eq.\ \ref{tau_s} should be replaced 
by the effective electron-phonon matrix element
\begin{equation}
    {\tilde F}_j({\bk,\bq}) = F_j({\bk,\bq}) -
		\left<F_j({\bk,\bq})\right>_{FS}.
    \label{ep_eff}
\end{equation}
Therefore, in what follows, Eq.\ \ref{tau_s} with $F_j$ replaced by
${\tilde F}_j$ will be assumed to be the correct expression for the phonon
relaxation rate. It should be noted that, for phonons propagating 
along high-symmetry directions, ${\tilde F}_j({\bk,\bq})$ differs 
from $F_j({\bk,\bq})$ only for longitudinal phonons.

The phonon lifetime in the normal state can be calculated from Eq.
(\ref{tau_s}) by putting the superconducting energy gap  equal to zero,
which gives
\begin{equation}
     \frac{1}{\tau_{{\bf q},j}} = \frac{8\omega_{{\bf q},j}^2
\tau_n}{\rho v_j^2}
     N_F \left[ \left\langle F^2_j(\bk,\bq) \right\rangle_{FS}
    - \left\langle F_j(\bk,\bq) \right\rangle_{FS}^2 \right],
     \label{tau_n}
\end{equation}
where $\tau_n$ is the electron lifetime in the normal state.

In some early work \protect\cite{tsu61,blo59} on
ultrasonic attenuation
the model Hamiltonian was formulated in terms of the potential energy
$V({\bf x})$, assumed
to be a function of the continuous electron position-coordinate ${\bf x}$,
and describing the interaction of the electron with the lattice and with
impurities. In this formulation, two problems arise that are eliminated
by a canonical transformation to a coordinate system fixed to the moving
lattice. The first is that the perturbation of the potential due to the
distorted lattice is not necessarily small.  This problem does not occur
in the tight-binding formulation because the perturbation is naturally
formulated in terms of a strain coordinate (i.e. the ratio of the
displacement to the phonon wavelength) rather than simply a displacement
coordinate [see Eq.\ (\ref{He-ph,plane})].  The second problem is that
it is simpler to work in a coordinate system in which the impurities
are static in the zero-order Hamiltonian so that the electron's
energy is conserved in the impurity
scattering process. In the tight-binding approach of this article the
impurities are static in the zero-order Hamiltonian.  For example,
potential scattering by impurities can be modelled by adding an extra
term $U_{\nu,{\bf n}_i} c_{\nu,{\bf n}_i,\sigma}^\dagger c_{\nu,{\bf n}_i,\sigma}$
to the Hamiltonian for the $\nu^{th}$ orbital on the impurity site
labelled by ${\bf n}_i$.  There is no dependence of this impurity
potential on the ion displacements, and the impurities are thus static
in this representation. Thus, our
approach has features qualitatively similar to those of some previous
 approaches
\protect\cite{tsu61,blo59}, even though the details of implementation
are different.

Before proceeding further, we show that our formulation reduces to the
traditional isotropic electron-stress-tensor formulation in an appropriate
limit. It is known \protect\cite{lan67} that for hexagonal crystals,
sound-wave propagation is isotropic with respect to rotations about the
c axis.  (This is not true for tetragonal crystals, such as
Sr$_2$RuO$_4$.)  Therefore, we expect
that the isotropic electron-stress-tensor model could give an appropriate
description of sound-wave attenuation in a two-dimensional hexagonal lattice,
particularly if the Fermi surface is taken to be a small circle about the
point ${\bf k}= 0$ so that the electron energy spectrum can be approximated by
that for free electrons, but with an effective mass.  For these reasons,
we consider for the moment a two-dimensional hexagonal lattice.
The quantity $F_j(\bk,\bq)$ defined in
Eq.\ (\ref{F}) can be written in the form
\begin{equation}
    F_j(\bk,\bq) = g \sum_{\alpha \beta}
        {\hat q}_\alpha f_{\alpha \beta}(\bk)e_{j\beta}
\end{equation}
where $\alpha$ and $\beta$ are summed over the $x$ and $y$ components of
two-dimensional vectors. Here
\begin{equation}
    f_{\alpha \beta}(\bk) = \sum_{\bf R} {\hat R}_\alpha {\hat R}_\beta
        \cos (\bk \cdot {\bf R})
\end{equation}
 and the sum over ${\bf R}$ is over three nearest-neighbor hexagonal
Bravais-lattice vectors that make angles of 120 degrees with each other.
Now, for a Fermi surface that is a small
circle surrounding the point $\bk = 0$, the approximation
$\cos(\bk \cdot {\bf R}) \approx
1 - \case 12 (\bk \cdot {\bf R})^2$ is valid.  This leads to the result
\begin{eqnarray}
{\tilde f}_{\alpha \beta}(\bk) & \equiv & f_{\alpha \beta}(\bk)
            -\langle f_{\alpha \beta}(\bk) \rangle_{FS}
            \nonumber \\
    & = & -\case 38 a^2(k_\alpha k_\beta -\case 12 k^2 \delta_{\alpha \beta})
\end{eqnarray}
where $a$ is the lattice constant.
This is precisely the form taken by the effective electron-phonon
interaction in the isotropic electron-stress-tensor model
used by many authors
\protect\cite{mor96,hir86,sch86,var86,graf00,tsu61,kad64,blo59,gra00,wu01},
and shows that our formulation of the theory of ultrasonic attenuation
is equivalent to theirs in an appropriate limit.

To prepare for a more detailed study of the ultrasonic attention as
observed in Ref.\ \protect\onlinecite{lup01} explicit expressions for
$F_j(\bk,\bq)$ [derived from Eq.\ (\ref{F})] are now given for our Sr$_2$RuO$_4$ model for the
case that both the phonon wave vector ${\bf q}$ and its polarization
vector ${\bf e}_j({\bf q})$ are in the basal plane.  Furthermore, a simple
model of isotropic phonons is assumed in which the polarization
vector is parallel to the wave vector for longitudinal phonons, and
perpendicular to the wave vector for transverse phonons, even when
the wave vector is not along a high symmetry direction.  The
direction of the phonon wave vector ${\bf q}$ in the basal plane is
characterized by the angle $\phi$ that it makes with the $x$ (i.e. $a$)
axis.

For longitudinal  phonons interacting with electrons in the $\gamma$
sheet of the Fermi surface via nearest-neighbor interactions
\begin{equation}
     F^\gamma_L(\bk,\bq) =
     g^\gamma [\cos^2 \phi \cos(k_x a)
             + \sin^2 \phi \cos(k_y a)].
     \label{lnn_gamma}
\end{equation}

For transverse $T_1$ phonons ($T_1$ phonons have their polarization as
well as their wave vector in the basal plane) interacting with
electrons in the $\gamma$ sheet of the Fermi surface via
nearest-neighbor interactions
\begin{equation}
     F^\gamma_{T1}(\bk,\bq) =
     g^\gamma \cos \phi \sin \phi
     [ \cos(k_x a) - \cos(k_y a)].
     \label{T_1nn_gamma}
\end{equation}

For longitudinal  phonons interacting with electrons in the $xz$
sheets of the Fermi surface via nearest-neighbor interactions
\begin{equation}
     F^{xz}_L(\bk,\bq) =
     g^{\alpha \beta} \cos^2 \phi \cos(k_F^{\alpha \beta}a),
     \label{lnn_xz}
\end{equation}
while for interactions with electrons in the $yz$ sheets of the Fermi surface
\begin{equation}
     F^{yz}_L(\bk,\bq) =
     g^{\alpha \beta} \sin^2\phi \cos(k_F^{\alpha \beta}a).
     \label{lnn_yz}
\end{equation}
In arriving at this result, this hybridization of the $xz$ and $yz$
bands has been neglected, as have interband transitions.  The bands
are thus one dimensional and characterized by the Fermi wave vector
$k_F^{\alpha \beta}$.

For transverse $T_1$  phonons interacting with electrons in the $xz$
sheets of the Fermi surface via nearest-neighbor interactions
\begin{equation}
     F^{xz}_{T1}(\bk,\bq) =
     -g^{\alpha\beta} \cos \phi \sin \phi \cos(k_F^{\alpha \beta}a),
     \label{T_1nn_xz}
\end{equation}
while for interactions with electrons in the $yz$ sheets of the Fermi surface
\begin{equation}
     F^{yz}_{T1}(\bk,\bq) =
     g^{\alpha \beta} \cos \phi \sin \phi \cos(k_F^{\alpha \beta}a).
     \label{T_1nn_yz}
\end{equation}

Now assume that the attenuation of longitudinal phonons is dominated by
their interaction with electrons  on the $ \alpha $ and $\beta $ sheets
of the Fermi surface as described by Eqs. (\ref{lnn_xz}) and
(\ref{lnn_yz}) (and that the interaction with electrons of the $ \gamma $ sheet can be neglected).  The attenuation will therefore be proportional to
\begin{eqnarray}
     \left\langle F^2_L \right\rangle_{FS}
    - \left\langle F_L \right\rangle_{FS}^2
     & = & p^{\alpha \beta}[g^{\alpha \beta}\cos(k^{\alpha \beta}_F)]^2
        \nonumber \\
    & \times & (\cos^4 \phi + \sin^4 \phi - p^{\alpha \beta})
    \label{attn_alpha_beta}
\end{eqnarray}
Here $ p^{\alpha \beta} $ is defined to be the fraction of the density of
states associated with the $xz$ (or the $yz$) band.  (The fraction of the
density of states associated with the $ \gamma $ band is called
$ p^\gamma$, so that
$ p^\gamma + 2p^{\alpha \beta} = 1$).

It is convenient to define the longitudinal anisotropy to be $ \eta_{l00}/ \eta_{110} $ where the mode viscosities of the L100 and L110 phonons, $\eta_{L100}$ and $\eta_{L110}$, respectively, are defined in the caption to Fig.\ \ref{Sr_expt}.  The experimental value of the longitudinal anisotropy is approximately 30 (see Fig.\ \ref{Sr_expt}).  The theoretical formula for the longitudinal anisotropy in the case that the attenuation is dominated by phonon interactions with electrons in the $xz$ and $yz$ bands is [from Eq.\ (\ref{attn_alpha_beta})]
\begin{equation}
    \frac{\eta_{L100}}{\eta_{L110}} =
    \frac{1 - p^{\alpha \beta}}{\case 12 - p^{\alpha \beta}}
    \label{la_alpha_beta}
\end{equation}
Using the experimentally determined value \protect\cite{mac96} $p^{\alpha \beta} = 0.21 $ gives a longitudinal anisotropy from Eq.\ (\ref{la_alpha_beta}) of 2.7. This value of the longitudinal anisotropy is too small by a factor of 10 to account for the experimentally observed value (which is 30).
 Thus interactions of longitudinal phonons with electrons in the $\alpha$
and $\beta$ sheets of the Fermi surface can not account for the
strong anisotropy  of the attenuation of
longitudinal phonons observed in Sr$_2$RuO$_4$ \protect\cite{lup01},
and these interactions must be dominated by other interactions.

Now suppose that the attenuation of longitudinal phonons is dominated by their interaction with electrons in the $\gamma$ band, and neglect the contribution to the attenuation from electrons in the $xz$ and $yz$ bands.  The longitudinal anisotropy in this case is given by
\begin{equation}
     \frac{\eta_{L100}}{\eta_{L110}}
         = \frac{ \langle \cos^2 (k_x a) \rangle_\gamma
            - p^\gamma \langle cos(k_x a) \rangle_\gamma}
             {\langle \{\case 12 [\cos(k_x a) + \cos(k_y a)]\}^2                            \rangle_\gamma
            - p^\gamma \langle cos(k_x a) \rangle_\gamma}.
     \label{la_gamma}
\end{equation}
Here the subscript $\gamma$ on the angular brackets (as in $\langle \rangle_\gamma$) indicates that average is over the $\gamma$ sheet of the Fermi surface only.  To begin with, for the purposes of obtaining a qualitative understanding of Eq.\ \ref{la_gamma}, neglect $ \langle cos(k_x a) \rangle_\gamma$ since numerical estimates show that it is small.
Now note that at the $X$ point of the Fermi surface, $k_x a = \pi$ and $k_y = 0$ so
the $\cos(k_x a) + \cos(k_y a) = 0$.  Because the $\gamma$ sheet of
the Fermi surface passes close to this point, the value of $\cos(k_x
a) + \cos(k_y a)$ will be small here and this will contribute to a
large value of the longitudinal anisotropy.  Furthermore, it should be
noted that the $X$ point is a saddle point of the electron energy
versus $\bk$ surface, so that the Fermi velocity $v_{\bf k}$ goes to
zero there.  This means that in carrying out the Fermi surface
average using Eq. (\ref{FSavg}), points close to the $X$ point will be
more heavily weighted, thus further enhancing thus magnitude of the
longitudinal anisotropy.  To obtain an explicit value for
longitudinal anisotropy, we make use of the tight-binding
approximation to $\varepsilon_{\bf k}$ given in Ref.\
\protect\onlinecite{maz97}, with the parameters given in that
article. Thus we use, for the electron energy in the $\gamma$ band,
$\epsilon_{\bf k} =
E_0 + 2t(\cos(k_xa) + \cos(k_ya)) + 4t^\prime \cos(k_xa)\cos(k_ya)$,
with parameters $(E_0 - E_F, t, t^\prime)$ = $(-0.4, -0.4, -0.12)$.
This allows Eq. (\ref{la_gamma}) to be evaluated numerically (including now the relatively small effect  of a nonzero $ \langle cos(k_x a) \rangle_\gamma$), giving the result
$(\eta_{L100}/\eta_{L110}) = 37$.  This is in reasonable agreement with the
experimentally determined value of approximately 30 (see Fig.\ \ref{Sr_expt}), considering the fact that no attempt was made to adjust the Fermi surface parameters to improve the agreement, and that the inclusion of a relatively small contribution from the less anisotropic attenuation due the the $xz$ and $yz$ electrons would also reduce the calculated longitudinal anisotropy.    The
conclusion is that electrons on the $\gamma$ sheet of the Fermi
surface give the dominant contribution to the ultrasonic attenuation,
and that the detailed Fermi surface structure for electrons on the
$\gamma$ sheet is important for understanding the large anisotropy in
the attenuation of the longitudinal sound waves.

So far, only the effect of the nearest-neighbor electron-phonon
interactions on the ultrasonic attenuation has been discussed.  Note
that for these interactions, the ultrasonic attenuation of the T100
waves is zero (see Eqs. (\ref{T_1nn_gamma}), (\ref{T_1nn_xz}), and
(\ref{T_1nn_yz}) for $\phi = 0$).  This is because transverse waves
propagating in the 100 direction do not stretch the nearest-neighbor
bonds.  Such waves do however stretch the next-nearest-neighbor
bonds, and attenuation of the T100 waves from the
next-nearest-neighbor electron-phonon interaction is to be expected.
This attenuation associated with $T_1$ basal-plane phonons interacting
with $\gamma$-sheet electrons, can be found in terms of the quantity

\begin{equation}
     F^{\prime \gamma}_{T1}(\bk,\bq) =
     g^{\prime \gamma} \cos(2\phi)
     \sin(k_x a) \sin(k_y a).
     \label{T_1nnn_gamma}
\end{equation}
The fact that the attenuation of the T100 phonons is about 1000 times
smaller than that of the L100 or T110 phonons is a good indication
that, in Sr$_2$RuO$_4$ the next-nearest-neighbor electron-phonon interactions are in
general less important than the near-neighbor interactions.

The formula
\begin{equation}
     \frac{\eta_{T100}}{\eta_{L100}} = \left(
         \frac{g^{\prime \gamma}}{g^{\gamma}} \right)^2
         \frac{\left\langle \sin^2(k_xa)\sin^2(k_ya) \right\rangle_{FS}}
             {\left\langle \cos^2(k_xa) \right\rangle_{FS}}
     \label{gprime}
\end{equation}
(from Eqs. (\ref{lnn_gamma}) and (\ref{T_1nnn_gamma})), together with the
experimentally measured values of the viscosities $\eta_{T100}$ and
$\eta_{L100}$ can be used to put an upper limit on the magnitude of
$g^{\prime \gamma}/g^\gamma$ of 0.041. This is an upper limit
because, as will be argued below, a different interaction could be
responsible for the T100 attenuation. This upper limit is surprisingly
small, given that the ratio of the next-nearest-neighbor to
nearest-neighbor hopping matrix elements for the $\gamma$ band as
estimated from the parameters assumed in Ref.\
\protect\onlinecite{maz97} is about 0.3.  There is, however, no {\it
a priori} reason why the sensitivity of a hopping matrix element to
bond stretching should be directly proportional to the magnitude of
the matrix element itself.

There is at present no experimental information on the attenuation of
$T_2$ phonons, i.e. by definition those transverse phonons with their
propagation vector ${\bf q}$ lying in the basal plane, and their
direction of polarization perpendicular to the basal plane.  The
attenuation obtained from a consideration of the body-diagonal
electron-phonon interaction described above is given in terms of
\begin{eqnarray}
   F^{\xi z,\eta z}_{T2}(\bk,\bq) & = &g^{\xi z,\eta z}
     \left[\cos \phi \sin(\case 12 k_x a) \cos(\case 12 k_y a)
      \right.     \nonumber   \\
     & + & \left. \sin \phi \cos(\case 12 k_x a) \sin(\case 12 k_y
a) \right]  \sin(\case 12 k_z c).
     \label{T2nn_xi_z}
\end{eqnarray}
Some factors of order unity have been absorbed into the definition of
$g^{\xi z,\eta z}$.  When squared and averaged over the Fermi
surface, this formula gives an attenuation independent of the
direction of ${\bf q}$ in the basal plane.

Similarly, the attenuation of $T_1$ phonons by the body-diagonal
electron-phonon interaction with the $xz$ and $yz$ bands is described
by
\begin{eqnarray}
     F^{\xi z,\eta z}_{T1}(\bk,\bq)  &=& \alpha g^{\xi z,\eta z}
     \cos(2\phi) \cos(\case 12 k_z c)\nonumber\\
   &&\times\sin(\case 12k_xa)\sin(\case 12 k_ya).
     \label{T_1nn_xi_z}
\end{eqnarray}
Here, $\alpha$ is a numerical constant of order unity.
The interaction of $T_1$ phonons by the body-diagonal electron-phonon
interaction with the $\gamma$-sheet electrons is described by an
identical formula, except we put $\alpha = 1$ and call the coupling
constant $g^{\prime \prime \gamma}$.

\section{Ultrasonic Attenuation in the Superconducting State}
\label{uss}

Now that a satisfactory model for the electron-phonon interaction has
been established, it is possible to proceed with confidence to an
interpretation of the ultrasonic attenuation measurements
\protect\cite{lup01} performed in the superconducting state of
Sr$_2$RuO$_4$. The basic idea is to use the principles described by
Moreno and Coleman \protect\cite{mor96} to gain information about the
location of the gap nodes in Sr$_2$RuO$_4$.  Evidence for the
existence of gap nodes, some of which is presented in Fig.\
\ref{Sr_expt} from Ref.\ \protect\onlinecite{lup01} has been
summarized in the introduction.

\begin{figure}
\centerline{\epsfig{file=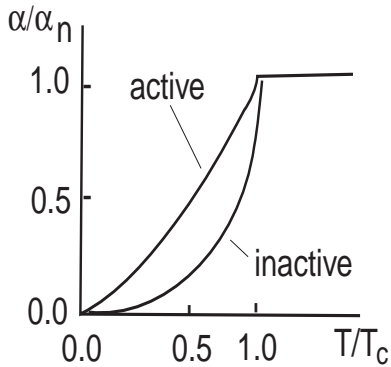,height=2in,width=2in}}
\vspace{10pt}
\caption{Qualitative behavior of the ultrasonic attenuation relative
to that in the normal state versus temperature relative to $T_c$.
The attenuation of sound for the case where only inactive nodes are
present grows more slowly by a factor of $T^2$ relative to that where
active nodes are present, at temperatures well below $T_c$.}
\label{act}
\end{figure}

At low temperatures in a superconductor with gap nodes, only
Bogoliubov quasiparticles in the neighborhood of the nodes are
thermally excited, and hence only these quasiparticles can interact
with the phonons to absorb them (and thus to attenuate a sound wave).
The lifetime of a phonon is given by Eq. (\ref{tau_s}).  Everything
depends on the behavior of the electron-phonon matrix
element ${\tilde F}_j(\bk,\bq)$ at the wave vectors ${\bf k}$ corresponding to the
nodes.  If this matrix element is nonzero at the nodes for a
particular phonon, then the phonon can interact well with the nodal
Bogoliubov quasiparticles, and the nodes are said to be active for
that particular phonon.  If, on the other hand, the matrix element is
zero at the nodes for a particular phonon, then the coupling of the
phonon to Bogoliubov quasiparticle precisely at the node is zero, and
grows as the distance from the node is increased.  In this case, the
nodes are said to be inactive for the phonon in question.  The
ultrasonic attenuation at temperatures well below the gap in the case
where only inactive nodes are present is proportional to $T^2$ times
the ultrasonic attenuation when active nodes are present
\protect\cite{mor96}, as illustrated in Fig.\ \ref{act}.

Suppose that the only nodes in the superconducting gap are vertical
line nodes in (110) planes.  Then it is clear from Eq.
(\ref{T_1nn_gamma}) that the electron-phonon matrix element for the T110
phonon is zero at the nodes
(since $k_x = \pm k_y$ for ${\bf k}$ in a (110) plane), and (110)
nodes are thus inactive for the T110 phonon.  For  T100 phonons, the
nearest-neighbor electron-phonon matrix element of Eq.
(\ref{T_1nn_gamma}) is zero for all ${\bf k}$ because $\phi = 0$.  Thus,
for the T100 phonon, its activity is determined by the
next-nearest-neighbor interaction given by Eq. (\ref{T_1nnn_gamma}), or Eq. (\ref{T_1nn_xi_z}),
which shows that the T100 phonon is active for nodes in (110) planes.
Similarly, Eq. (\ref{lnn_gamma}) can be used to show that the electron-phonon matrix element ${\tilde F}$
is nonzero at nodes in (110) planes for both L100 ($\phi = 0$) and
L110 ($\phi = \pi/4$) phonons, showing that both of these modes are
active.  Thus, if the only nodes in the gap are those in (110)
planes, the attenuation of the T110 sound wave would increase
significantly more slowly on increasing the temperature from zero
than that of the T100, L100 or L110 sound waves.  From Fig.\
\ref{Sr_expt} this is clearly not the case, thus ruling out states
that have nodes only in (110) planes.  In particular this rules out a
superconducting spin-singlet state of $k_x^2 - k_y^2$ symmetry, as
well as a spin-triplet $f$-state with ${\bf d} = {\bf \hat{z}}(k_x +
ik_y)(k_x^2 - k_y^2)$.

Next consider the case that the only nodes in the superconducting gap
are vertical line nodes in the (100) planes.  Arguments similar to those given
above show that these (100) nodes are active for L100, L110, and T110
sound waves, but inactive for T100 sound waves.  In this case the
attenuation of the T100 sound wave should increase markedly more slowly
with temperature at low temperatures than that of the other three
sound wave types, and this is not the case.  This rules out states
that have only (100) line nodes, such as the $f$-wave state with
${\bf d} = {\bf \hat{z}}(k_x + ik_y)k_x k_y$.

The arguments of the previous paragraphs have ignored the effects of
the $\alpha$ and $\beta$ sheets of the Fermi surface. Our
implementation of our model is inaccurate for the behaviour of the
electron-phonon matrix element at (110) nodes for the $\alpha$ and
$\beta$ sheets of the Fermi surface because we have not taken into
account properly the hybridization of the $xz$ and $yz$ bands, which
is important in the [110] directions.  A more detailed discussion
could be given to remedy this deficiency. This will not be necessary,
however, as it will be seen that the more general and powerful
symmetry arguments of the following section confirm the conclusions
reached above.

It should be noted that the electron-phonon interaction developed in
this article gives results that are significantly different from
those obtained when  the electron-phonon interaction is formulated in
terms of an isotropic electron stress tensor (e.g. see Refs.\
\protect\onlinecite{mor96,gra00,wu01}).  This can seen, for example,
in the top panel of Fig. 2 of Ref.\ \protect\onlinecite{wu01}, where
a model with gap nodes in (110) planes is analyzed, and the L100 mode
shows behavior characteristic of interaction with inactive nodes,
since its calculated attenuation grows much more slowly as the
temperature is increased from zero than the calculated attenuation of
the T100 mode.  As noted in Ref.\ \protect\onlinecite{wu01}, the
intensity of the coupling of the L100 phonon to the electrons in the
isotropic electron-stress-tensor model is proportional to the factor
$(\hat{k}_x^2 - \frac{1}{2})^2$, which is zero in (110) planes,
making the (110) nodes inactive.  This inactivity of the (110) nodes
for the L100 phonon is an ``accidental'' inactivity, i.e. it is a
particularity of the isotropic electron-stress-tensor electron-phonon
interaction which would not be present in a more general model and
(as demonstrated in the follow section) is not required by symmetry.
In fact, as shown above, the (110) nodes are active for L100 phonons
for the electron-phonon interaction developed in this article.  Thus
it seems clear that the isotropic electron-stress-tensor model should not be
used in interpretations of ultrasonic attenuation experiments which
aim at locating the positions of the gap nodes in an unconventional
superconductor.

Finally note that the electron-phonon matrix element for the
attenuation of transverse $T_2$ phonons in Sr$_2$RuO$_4$ (i.e. those
that have a propagation vector ${\bf q}$ in the basal plane and a
polarization perpendicular to the basal plane) is given by Eq.
(\ref{T2nn_xi_z}).  The factor $\sin(k_z c/2)$ contained in
this expression means that this matrix element is zero for horizonal
nodes lying in the plane
$k_z = 0$ or in on the $k_z = 2\pi/c$ surface of the Brillouin zone.
Thus, horizontal nodes at $k_z = 0$ or at $k_z = 2\pi/c$ are inactive
for $T_2$ phonons,
and the comparison of the attenuation for $T_2$ phonons with that for
other active phonons can be used as a test for such nodes.
Unfortunately, it is more difficult to find a definitive test for
horizontal nodes that might appear at
$k_z = \pi/c$, such as in the proposal of Ref.\
\protect\onlinecite{zhi01}, since there does not appear to be any
phonon mode that is inactive for such nodes.  This is related to the
fact that such nodes are not required by the symmetry of the order
parameter, but are accidental.

So far we have investigated only the effects of the relatively large
differences in the temperature dependence of the attenuation that are
expected to occur as a result of the differences between active and
inactive nodes.  Such differences do not appear to occur in the
experimental results currently available for Sr$_2$RuO$_4$.  There
is however an intriguing smaller difference in temperature dependence
noted by Lupien et al. \protect\cite{lup01}  This is that the ratio of
the attenuation of the T100 mode in the superconducting state to its
normal-state value is somewhat larger at low temperatures than that
of the other measured sound wave modes (see Fig.\ \ref{Sr_expt}).
This attenuation is much smaller than that of the other measured
sound wave modes, and was accounted for above in terms of a
second-neighbor intraplanar electron-phonon interaction.  This is not
the only possibility, however.  It is possible that the interplanar
body-diagonal electron-phonon interaction could be the most important
for this mode.  This interaction has the characteristic feature that
the squared matrix element contains the factor $\cos^2(k_z c/2)$.
If there are horizontal nodes close to either the $k_z = 0$ or $k_z =
2\pi/c$ planes where this factor has its maximum, then this would
give the T100 sound wave modes a boost in attenuation at low
temperatures relative to the other observed nodes, in agreement with
experiment. 

In summary, the results of ultrasonic attenuation experiments
\protect\cite{lup01}, when looked at in the light of the above
theoretical considerations, provide  evidence against vertical 
line nodes in either [100] or
[110] planes.  Also, the experiments are consistent with
horizontal line nodes.  Finally, as to the positions of the horizontal
line nodes, positions close to either the $k_z = 0$ or $k_z = 2\pi/c$
planes appear to be favored.

\section{Crystallographic Symmetry, the Electron-Phonon Interaction, and
Active Nodes}
\label{symmetry_arguments}

In the last section an explicit model of the electron-phonon
interaction was used to deduce the activity of nodes of different
types relative to a given sound wave mode.  While there is reason to
believe that the model for the electron-phonon interaction is a
relatively good one, since it accounts well for a number of unusual
features of the sound attenuation in the normal state of
Sr$_2$RuO$_4$, it is nevertheless of interest to develop an approach
that is not dependent on the details of the particular model.  Such
an approach, which exploits the crystallographic symmetry, is indeed
possible, and will now be sketched.  Such symmetry arguments will be
particularly useful in cases where the electronic structure may not
be well known, or is sufficiently complicated that a detailed model
is difficult to develop.

In this section, the Hamiltonian describing the electron-phonon
interaction will be taken to be that of Eq. (\ref{He-ph}), where the
matrix element $g_{{\bf k}+\frac{1}{2}{\bf q,q},j}^\nu $ will be
assumed to be completely general, and subject only to the
restrictions imposed by symmetry.  For example, from the properties
of the Hermiticity of the Hamiltonian, its time-reversal invariance,
and its invariance with respect to spatial inversion in a ruthenium
ion position, it follows that $g_{{\bf k}+\frac{1}{2}{\bf q,q},j}^\nu
$ is pure imaginary, is an even function of its argument ${\bf k} +
\frac{1}{2}{\bf q}$, and is an odd function of its argument ${\bf q}$.

Symmetry arguments can be used to show that, for a transverse phonon
with wave vector ${\bf q}$ along the [110] direction, and its
polarization $j$ in the basal plane, perpendicular to ${\bf q}$,
vertical lines nodes in (110) planes are inactive.  Suppose that
${\bf k}$ has its basal-plane component in  the [110] direction. Then
under a reflection in a plane normal to the basal plane, and
containing ${\bf q}$,
$c_{\nu,{\bf k},\sigma}$ must have a definite parity (i.e. either
$c_{\nu,{\bf k},\sigma} \rightarrow + c_{\nu,{\bf k},\sigma}$ or
$c_{\nu,{\bf k},\sigma} \rightarrow - c_{\nu,{\bf k},\sigma}$).  Also,
$A_{{\bf q},j} \rightarrow - A_{{\bf q},j}$. From these properties it
follows that $g_{{\bf k}+\frac{1}{2}{\bf q,q},j}^\nu = 0$ for all
${\bf k}$ lying in the same (110) plane that contains ${\bf q}$.  A
related argument shows that for ${\bf
k}$ lying in the plane perpendicular to ${\bf q}$, $g_{{\bf
k,q},j}^\nu = 0$.  For this latter argument, it is necessary to
assume that $|{\bf q}| \ll |{\bf k}|$, so that $g_{{\bf
k}+\frac{1}{2}{\bf q,q},j}^\nu$ can be replaced by $g_{{\bf
k,q},j}^\nu$.  Thus, for a T110 sound wave, vertical line nodes in
(110) planes are inactive.
Furthermore, there are no symmetry arguments that show that (110)
nodes are inactive for T100, L100 or L110 sound wave, and they must
therefore, in general, be active.

In a similar way it can be shown that symmetry requires that (100)
vertical nodes are inactive for T100 sound waves, while these nodes
are in general active for T110, L100 and L110 sound waves.

Making use of the fact that the basal plane is a plane of reflection
symmetry, one can also show that line nodes in the plane
$k_z = 0$, and in the plane $k_z = 2\pi/c$ are inactive for $T_2$ sound
waves having ${\bf q}$ in the basal plane and polarized perpendicular
to the basal plane.

In the long-wavelength limit $|\bq|\ll|\bk|$, the symmetry arguments
given above can be formulated in a particularly simple form. In this limit,
the function ${\tilde F}^\nu_j(\bk,\bq)$  determining the
symmetry of the electron-phonon interaction [and which can be obtained from
from Eqs. (\ref{ep_eff}) and (\ref{F})]can be written as
\begin{equation}
 {\tilde F}^\nu_j(\bk,\bq)=g^\nu \sum_{\alpha,\beta = x,y,z}
    f^\nu_{\alpha\beta}(\bk)\hat q_\alpha e_{j,\beta}.
\end{equation}
(Note that, in the isotropic electron-stress-tensor model,
$f_{\alpha\beta}(\bk)=\hat k_\alpha\hat k_\beta-(1/d)\delta_{\alpha\beta}$, where
$d$ is the dimensionality of the system; this isotropic stress-tensor expression is not used here.)
If, for a given $\bk$, there is a symmetry operation from the crystallographic
point group which leaves $\bk$ invariant, but changes sign of
$\hat q_\alpha e_{j,\beta}$, then the matrix element
$f^\nu_{\alpha\beta}(\bk)$ vanishes (one should also use the fact that in symmetry operations, ${\bf e}_j$ and ${\bf q}$ transform like vectors, and
${\bf e}_j(-\bq)={\bf e}_j(\bq)$). In particular,
one of the consequences is that there are no zeros of
${\tilde F}^\nu_j(\bk,\bq)$
required by symmetry for all longitudinal waves, so the nodes are always
active for this polarization.
Also, it is clear that a symmetry-imposed line zeros for transverse waves
can be present only if $\bk$ is in a high-symmetry plane.

Interband transitions of electronic quasiparticles have been ignored
in our discussion.  Such transitions are expected to play a role only
when two bands cross, and since there are relatively few
quasiparticles associated with such points, their effects in general
should not be important.  However, if a case is encountered where
interband transitions play a significant role they will have to be
investigated carefully, since the symmetry properties of the
electron-phonon matrix element for such transitions are different from
those for intraband transitions.

As an application of these ideas to a material with a relatively
complicated Fermi surface \protect\cite{tai87,ogu87}, consider
ultrasonic attenuation in UPt$_3$, \protect\cite{shi86,ell96} which
is believed to give evidence for the existence of basal-plane line
nodes in the superconducting gap.  We examine the consequences of the
conjecture that the order parameter is characterized by a
singlet-state gap $\Delta(k_x, k_y, k_z)$  transforming in accordance
with the E$_{2g}$ representation of the point group D$_{6h}$, or by
the triplet-state order parameter $d_z(k_x, k_y, k_z)$ which
transforms according to E$_{2u}$.
(These are two of the most commonly assumed candidates for the order
parameter, e.g. see Ref.\ \protect\onlinecite{gra00}.)  First note
that from their symmetry classification, both of these order
parameters change sign under reflection in the basal plane. (The
space group of UPt$_3$ is P6$_3$/mmc, which has a $\sigma_z$
reflection plane.)  Hence,
\begin{eqnarray}
    \Delta(k_x, k_y, k_z) & = & -\Delta(k_x, k_y, -k_z) \nonumber \\
    d_z(k_x, k_y, k_z) & = & -d_z(k_x, k_y, -k_z).
    \label{Delta}
\end{eqnarray}
The constraints of the this equation require the order parameters to
be zero at $k_z = 0$, i.e. there will be a line of nodes wherever a
sheet of the Fermi surface cuts the basal plane.  (The Fermi surface
consists of several sheets \protect\cite{tai87,ogu87}, with each
sheet having its own gap, but all gaps are expected to have the same
symmetry, and to  satisfy Eq. (\ref{Delta}).)  Note that Eq.
(\ref{Delta}) also requires the order parameters to be zero at surfaces
of the hexagonal Brillouin zone at $k_z = \pm \pi/c$.  This is
because $(k_x, k_y, +\pi/c)$ and  $(k_x, k_y, -\pi/c)$ are equivalent
points.  There are two toroidal sheets of the Fermi surface that cut
the surface $k_z = \pi/c$ of the Brillouin zone,
\protect\cite{tai87,ogu87} and hence give line nodes there.  The
presence of these Brillouin-zone surface line nodes, associated by
symmetry with those in the basal plane, does not seem to have been
previously noticed.

Because the basal plane of UPt$_3$ is a plane of reflection symmetry,
symmetry arguments similar to those given above show that both
basal-plane line nodes and $k_z = \pi/c$ Brillouin-zone-surface line
nodes are inactive for $T_2$ sound waves with ${\bf q} \parallel {\bf
a}$ and polarization parallel to ${\bf c}$, but active for $T_1$ sound
waves with ${\bf q} \parallel {\bf a}$ and polarization parallel to
${\bf b}$. The temperature dependence of the ultrasonic attenuation
data \protect\cite{shi86,ell96} shows that the nodes are active for
$T_1$ sound waves, and inactive for $T_2$ sound waves, giving clear
evidence of the existence of basal plane and $k_z = \pi/c$ Brillouin-zone-surface line nodes.  These comments extend previous discussions to
include Brillouin-zone-surface line nodes, and also ensure that the
analysis remains valid for the complex Fermi surface geometry that
actually occurs in UPt$_3$.

\section{conclusions}
The electron-phonon interaction in Sr$_2$RuO$_4$ has a strong anisotropy
that is highly unusual.  The unusual nature of this interaction can
be attributed to certain structural properties of Sr$_2$RuO$_4$,
these being that the structure is a layered one with a relatively large
distance between the RuO$_4$ layers, and that the Ru ions in a layer
form a square lattice with the largest electron-phonon interaction
being between nearest-neighbor Ru ions.
A detailed model electron-phonon interaction based on the idea of a
tight-binding Hamiltonian with hopping matrix elements that depend on
the ion displacements does a good job of quantitatively accounting
for the huge anisotropy observed \protect\cite{lup01} in ultrasonic
attenuation in the normal state of Sr$_2$RuO$_4$.  The dominant
contribution to the attenuation comes from the interaction of phonons
with electrons in the $\gamma $ band. The attenuation of transverse
sound waves propagating in the [100] direction and having their
polarization in the basal plane is exceptionally small because such
waves do not stretch the nearest-neighbor bond lengths, and hence
have no nearest-neighbor electron-phonon interaction. The strong
anisotropy for the propagation of longitudinal waves in the basal
plane can be related to the fact that the $\gamma $ sheet of the
two-dimensional Fermi surface describes a large circle passing close
to the $X$ point of the Brillouin zone. An analysis of the ultrasonic
attenuation data \protect\cite{lup01} in the superconducting state in
terms of our model electron-phonon interaction rules out the
possibilities that the only nodes in the superconducting gap are
vertical lines nodes in (100) planes, or vertical line nodes in (110)
planes.  The experiments performed so far  are consistent with the
existence of horizontal line nodes such as those proposed in
Refs.\ \protect\onlinecite{has00} and \protect\onlinecite{zhi01}. 
With respect to the positions of horizontal line nodes,
positions close to the
planes $k_z = 0$ or $k_z = 2\pi/c$ appear to be favored, but a more
detailed analysis is need to confirm this.

A general method (based on crystallographic symmetry arguments) of
determining the existence of inactive nodes for a given sound wave
mode is developed.  This method is useful even when it is not
possible to develop a detailed microscopic model for the electron-phonon
interaction.

As a by-product of our theory, we also propose that the ultrasonic
attenuation data in UPt$_3$ can be interpreted in favor of the existence
of horizontal line nodes in the plane $k_z=\pm\pi/c$, as well as in the
plane $k_z = 0$.

Our results, both for the anisotropy of the ultrasonic attenuation in
the normal state, and for questions related to the positions of gap
nodes in the superconducting state, are different from those obtained
when using the traditional method of modeling the electron-phonon
interaction is terms of an isotropic electron stress tensor.  In particular,
the isotropic electron-stress-tensor model  makes certain superconducting gap
nodes accidentally inactive, when they are in fact active.

\section*{acknowledgements}
We thank L. Taillefer, C. Lupien, and A. MacFarlane for stimulating
discussions, and C. Lupien for
providing us with Fig.\ \ref{Sr_expt}. The hospitality 
of Efim Kats and the Theory Group of the
Institut Laue Langevin, where some of this work was done, is much appreciated,  as is 
the support of the Canadian Institute for Advanced Research 
and of the Natural Sciences and Engineering Research Council of Canada.

\section*{appendix: symmetry and universality of ultrasonic attenuation}

One of the remarkable features of unconventional superconductors is
that some of their transport coefficients attain ``universal'' (i.e.
independent of the disorder concentration) values at sufficiently low
temperatures. In particular, universality was first predicted for
 electrical conductivity of $d$-wave
superconductors in Ref.\
\protect\onlinecite{lee93} (see also Ref.\ \protect\onlinecite{sun95}),
and for thermal conductivity in Refs.\ \protect\onlinecite{nor96,gra96}
and observed
experimentally in thermal conductivity in Ref.\ \protect\onlinecite{tai97}.
In this Appendix, we study the low-temperature behaviour of the ultrasonic
attenuation in unconventional superconductors, using the model formulated
in Sec. \ref{model_formulation}.

In the absence of the vertex corrections, the ultrasonic attenuation
coefficient in the hydrodynamic approximation is given by
\begin{equation}
\label{alpha_general}
   \frac{\alpha_j(\bq,T)}{\alpha_j(\bq,T_c)}=
   \frac{1}{\tau_n}\int_0^\infty
   d\epsilon \left(-\frac{\partial f}{\partial \epsilon}\right)
   \frac{A_j(\bq,\epsilon)}{\re\tveps(\epsilon)},
\end{equation}
where $j=L,T_1,T_2$ labels the phonon polarization, and
\begin{eqnarray}
   &&A_j(\bq,\epsilon)=\frac{1}{2\im\tveps(\epsilon)}
   \left\langle {\tilde F}_j^2(\bk,\bq)\right\rangle_{FS}^{-1}\nonumber\\
   &&\times \left\langle {\tilde F}_j^2(\bk,\bq)
  \re\frac{|\tveps(\epsilon)|^2+\tveps^2(\epsilon)-2|\Delta_\bk|^2}{\sqrt{\tveps^2(\epsilon)-
   |\Delta_\bk|^2}} \right\rangle_{FS}.
\end{eqnarray}
The temperature-independent ultrasonic attenuation in the normal
state is given by
\begin{equation}
\label{alpha_normal}
   \alpha_j(\bq,T_c)=\frac{8\omega^2_{\bq, j}}{\rho v^3_j}
   N_F\tau_n\left\langle
   {\tilde F}_j^2(\bk,\bq)\right\rangle_{FS}.
\end{equation}
The expression (\ref{alpha_general}) is similar to that given in Ref.
\protect\onlinecite{hir86}, the only difference being in the angular
dependence of
the electron-phonon interaction (see also Ref.\ \protect\onlinecite{graf00}).
To reproduce the results
obtained with the help of the isotropic electron stress tensor, one should replace
${\tilde F}_j(\bk,\bq)\to (\hat\bk\cdot\hat\bq)(\hat\bk\cdot{\bf e}_j(\bq))-(1/d)
(\hat\bq\cdot{\bf e}_j(\bq))$.

The function $\tveps(\epsilon)$ describes the self-consistent
renormalization of
quasiparticle energy due to impurity scattering and satisfies the equation
\begin{equation}
\label{selfcon}
   \tveps=\epsilon+\frac{i}{2\tau_n}\frac{g(\tveps)}{\cos^2\delta_0+
   g^2(\tveps)\sin^2\delta_0},
\end{equation}
where $g(\epsilon)$ is the retarded Green's function $G$ at coinciding
points (see Ref.\ \protect\onlinecite{Book} for a review of the effects of
disorder in unconventional superconductors). Assuming electron-hole symmetry,
\begin{equation}
   g(\epsilon)=\left\langle\frac{\epsilon}{\sqrt{\epsilon^2-|\Delta_\bk|^2}}
   \right\rangle_{FS}.
\end{equation}
In particular, for quasi-2D $d$-wave order parameters
$\Delta_\bk=\Delta_0\cos 2\varphi$ ($d_{x^2-y^2}$ symmetry) or
$\Delta_\bk=\Delta_0\sin 2\varphi$ ($d_{xy}$ symmetry) and a
cylindrical Fermi surface,
$g(\epsilon)=(2/\pi)K(\Delta_0/\epsilon)$,
where $K(x)$ is the complete elliptic integral.
The function $\tveps(\epsilon)$ determines, for example, the inverse mean
free time of quasiparticles:
\begin{equation}
 \label{scatter_rate}
   \frac{1}{\tau(\epsilon)}=2\im\tveps(\epsilon),
\end{equation}
and the disorder-averaged quasiparticle density of states (DoS):
\begin{equation}
   N(\epsilon)=N_F\re g(\tveps(\epsilon)).
\end{equation}

The behaviour of the attenuation coefficient strongly depends on the
value of the
phase shift $\delta_0$. Here we consider two limiting cases of weak impurities
(Born limit), when $\delta_0\to 0$, and strong impurities (unitary limit), when
$\delta_0\to\pi/2$.

The solution of Eq. (\ref{selfcon}) at zero energy is purely imaginary in both
cases: $\tveps(\epsilon=0)=i\Gamma_0$, where
\begin{equation}
   \Gamma_0=\Delta_0\exp(-\pi\Delta_0\tau_n)
\end{equation}
in the Born limit, and
\begin{equation}
   \Gamma_0=\Delta_0\sqrt{\frac{\pi}{2\Delta_0\tau_n\ln\Delta_0\tau_n}}
\end{equation}
in the unitary limit. The zero-energy scattering rate $\Gamma_0$ determines
the crossover energy scale separating two qualitatively different types of
the behaviour of the observable quantities.
If the typical energy of excitations (temperature) is greater than
$\Gamma_0$, then one can neglect the self-consistent energy renormalization
and use the quasiparticle Boltzmann equation for calculating the kinetic
properties in the superconducting state (for the application of this approach
to unconventional superconductors, see, e.g. Ref.\ \protect\onlinecite{arfi88}).
In contrast, if the typical energy is smaller than
$\Gamma_0$, then the self-consistency effects become important, and
the quasiparticle Boltzmann equation is not applicable.
In the former case, the imaginary part of $\tveps$ is small
compared to $\re\tveps\to\epsilon$, and we obtain
\begin{equation}
   \label{alpha_highT}
   \frac{\alpha_j(\bq,T)}{\alpha_j(\bq,T_c)}=\frac{1}{\tau_n}\int_0^\infty
    d\epsilon \left(-\frac{\partial f}{\partial \epsilon}\right)
    \frac{A_j(\bq,\epsilon)}{\epsilon},
    \label{attn1}
\end{equation}
where
\begin{equation}
   A_j(\bq,\epsilon)=2\tau(\epsilon)
   \frac{\left\langle {\tilde F}_j^2(\bk,\bq)\re\sqrt{\epsilon^2-|\Delta_\bk||^2}
   \right\rangle_{FS}}{\left\langle
   {\tilde F}_j^2(\bk,\bq)\right\rangle_{FS}}.
    \label{attn2}
\end{equation}
The expression (\ref{alpha_highT}) is equivalent to Eq. (\ref{tau_s}) for
isotropic impurity scattering.
It should be noted that the quasiparticle scattering rate
(\ref{scatter_rate}) in the Born limit decreases with energy in the
superconducting state, and at some temperature $T^*\simeq (\omega\tau_n)T_c$
the applicability condition of the hydrodynamic approximation is
violated. It turns out, however, that in real experimental conditions
for Sr$_2$RuO$_4$, the crossover temperature $T^*$ is so small that we neglect
this complication here.

In the low-temperature regime, we replace $\im\tveps$ by
$\Gamma_0$, take the limit $\re\tveps\to 0$, calculate the integral
over energy,
and obtain
\begin{equation}
\label{unitary_lowT}
\frac{\alpha_j(\bq,T)}{\alpha_j(\bq,T_c)}=\frac{1}{2\tau_n}
   \frac{\left\langle {\tilde F}_j^2(\bk,\bq)\displaystyle
   \frac{\Gamma_0^2}{(\Gamma_0^2+|\Delta_\bk|^2)^{3/2}}
   \right\rangle_{FS}}{\left\langle
{\tilde F}_j^2(\bk,\bq)\right\rangle_{FS}}.
\end{equation}
We see that, at $T<\Gamma_0$, the ultrasonic attenuation does not depend on
temperature.

In the Born limit, for typical disorder concentrations in cuprates
and ruthenates,
$\Gamma_0$ turns out to be exponentially small compared to $T_c$ (in
particular,
in the experimental conditions of Ref. \protect\onlinecite{lup01},
$\Delta_0\tau_n\sim l_{n,ab}/\xi_{0,ab}\sim 10$,
so that $\Gamma_0/T_c < 10^{-3}$). For this
reason, only the ``high-temperature'' limit is relevant for Born impurities.

For unitary impurities, however, $\Gamma_0$ can be as large as $0.1T_c$, which
means that, on one hand, there might be no clear power-law behaviour
of $\alpha(T)$ at $\Gamma_0<T\ll T_c$ (it was pointed out, e.g. in
Ref.\ \protect\onlinecite{hir86}), and, on the other hand,
the low-temperature regime can be observable. The contributions to the
ultrasonic attenuation from active and inactive nodes can be easily separated
because of their different dependences on the impurity concentration.
For active line nodes at a cylindrical Fermi surface,
\begin{equation}
  \label{average_act}
   \left\langle {\tilde F}_j^2(\bk,\bq)
   \frac{\Gamma_0^2}{(\Gamma_0^2+|\Delta_\bk|^2)^{3/2}}
   \right\rangle_{FS}=\frac{F_0^2}{\pi}\frac{1}{\Delta_0},
\end{equation}
where $F_0$ is the value of ${\tilde F}_j(\bk,\bq)$ at the line of nodes. Therefore,
as seen from Eqs. (\ref{alpha_normal}) and (\ref{unitary_lowT}),
the ultrasonic attenuation does not depend on the impurity concentration.
In contrast, for inactive line nodes,
\begin{equation}
 \label{average_inact}
   \left\langle {\tilde F}_j^2(\bk,\bq)
   \frac{\Gamma_0^2}{(\Gamma_0^2+|\Delta_\bk|^2)^{3/2}}
   \right\rangle_{FS}=\frac{F'_0{}^2}{\pi}\frac{\Gamma_0^2}{\Delta_0^3}
   \ln\frac{\Delta_0}{\Gamma_0},
\end{equation}
where $F'_0$ is the value of the transverse derivative of ${\tilde F}_j(\bk,\bq)$ at the line of
nodes. Therefore, in this case, the ultrasonic attenuation does depend on the impurity concentration.
Comparing Eqs. (\ref{average_act}) and (\ref{average_inact}), we see that the contribution
from inactive nodes is typically much smaller than
that from active ones, and one can make a conclusion that if, for a given
polarization and propagation, there are active nodes present, the attenuation
coefficient is universal at low temperatures.
In particular, the attenuation of longitudinal waves should always be universal.
On the other hand, for example, the attenuation of the in-plane $T_2$ waves for the order
parameters with horizontal line nodes at $k_z=0$ cannot be universal, because such nodes
are inactive.
Also, for a $f$-wave order parameter
${\bf d}({\bf k})\propto{\bf{\hat{z}}}(k_x+ik_y)k_xk_y$, the attenuation
coefficient of the $T110$ phonons is universal, whereas that of the $T100$ phonons is not.
For the order parameter ${\bf d}({\bf k})\propto{\bf{\hat{z}}}(k_x+ik_y)(k^2_x-k^2_y)$
the situation is opposite: the $T100$ attenuation is universal, but $T110$ is
not. It should be noted that the calculations based on the isotropic electron-stress-tensor model give different results \cite{graf00}.

\end{document}